\documentclass{article}
\usepackage{xcolor}
\usepackage{booktabs} 
\usepackage{cite}
\usepackage{algorithmic}
\usepackage[ruled,linesnumbered]{algorithm2e} 
\usepackage{mdwmath}
\usepackage{mdwtab}

\usepackage{graphicx}
\usepackage{lipsum}
\usepackage{multicol}
\usepackage{multirow}
\usepackage{makecell} 
\usepackage{mathtools}
\usepackage{subcaption}
\usepackage[bottom]{footmisc}
\usepackage{balance}

\begin{document}
\title{Hierarchical Dynamic Loop \mbox{Self-Scheduling} \\on Distributed-Memory Systems Using an MPI+MPI Approach}

\author{Ahmed Eleliemy and Florina M. Ciorba \\
	Department of Mathematics and Computer Science\\University of Basel, Switzerland\\
		\{firstname.lastname\}@unibas.ch
}

%
%
\date{}
\maketitle
\newpage
\tableofcontents
\newpage
\begin{abstract}

\mbox{Computationally-intensive} loops are the primary source of parallelism in scientific applications.
Such loops are often irregular and a balanced execution of their loop iterations is critical for achieving high performance.
However, several factors may lead to an imbalanced load execution, such as problem characteristics, algorithmic, and systemic variations.
Dynamic loop \mbox{self-scheduling}~(DLS) techniques are devised to mitigate these factors, and consequently, improve application performance. 
On \mbox{distributed-memory} systems, DLS techniques can be implemented using a hierarchical \mbox{master-worker} execution model and are, therefore, called hierarchical DLS techniques.
These techniques \mbox{self-schedule} loop iterations at two levels of hardware parallelism: across and within compute nodes. 
Hybrid programming approaches that combine the message passing interface~(MPI) with open \mbox{multi-processing}~(OpenMP) dominate the implementation of hierarchical DLS techniques.
The \mbox{MPI-3} standard includes the feature of sharing memory regions among MPI processes. 
This feature introduced the MPI+MPI approach that simplifies the implementation of parallel scientific applications.
The present work designs and implements hierarchical DLS techniques by exploiting the MPI+MPI approach.
Four \mbox{well-known} DLS techniques are considered in the evaluation proposed herein. 
The results indicate certain performance advantages of the proposed approach compared to the hybrid MPI+OpenMP approach.\\


\end{abstract}
\newpage
\section{Introduction}
\label{sec:intro}
%

Today, multi- and many-core processors are at the basis of modern \mbox{large-scale} \mbox{distributed-memory} systems.
The message passing interface~(MPI)~\cite{MPIForum} is the most dominant programming approach for scientific applications that execute on \mbox{large-scale} \mbox{distributed-memory} systems.
MPI provides portable performance on most of the modern \mbox{distributed-memory} systems. 
However, the increasing number of computing cores within a single processor creates specific performance challenges for MPI.
For instance, mapping MPI processes to individual cores results in limited \mbox{sub-problem} sizes per MPI process; the memory capacity per core does not grow as fast as the growth of the number of cores within a processor~\cite{MPIShm1}.
Also, exchanging data between two MPI processes that execute on the same processor may pose another performance challenge;  
this data exchange stresses the memory subsystems by performing \mbox{memory-to-memory} copy operations which may degrade applications performance~\cite{MPIShm2}.
These challenges motivated the intensive use of hybrid parallel programming in scientific applications. 
The hybrid programming approaches combine distributed-memory and shared-memory programming approaches. 
For instance, the combination of MPI with OpenMP~\cite{OpenMP}, known as \mbox{MPI+OpenMP}, is widely used in several scientific applications~\cite{MPIOpenMP1,MPIOpenMP2,MPIOpenMP3}. 
However, exploiting two parallel programming models and managing parallelism by two different runtime systems within the same application may lead to complex applications codes. 
The code complexity may pose performance and maintainability challenges.    

The MPI forum added, in the \mbox{MPI-3} standard, the feature of sharing memory regions between MPI processes that reside on a \mbox{shared-memory} system~\cite{MPIForum}.
This introduced a new parallel programming approach called \emph{MPI+MPI approach}~\cite{MPIShm1}.
The new approach can be seen as a technological advancement that may result in enhancing applications' performance and reducing applications' code complexity.
The present work focuses on exploiting the \mbox{MPI+MPI} approach in designing and implementing efficient hierarchical dynamic loop self-scheduling~(DLS) techniques.

Executing scientific applications on \mbox{multi-processor} systems requires all processing elements to reach nearly equal finishing times.
DLS techniques aim to balance loop iterations' execution across all processing elements by mitigating the effects of several factors, such as problem characteristic, algorithmic and systemic variations, which may hinder such a balanced execution~\cite{AWF,LIB}.

DLS techniques typically employ a \mbox{master-worker} execution model~\cite{LBtool,DLBLtool,processorgroups}.
This model includes a processing entity called master that is responsible for calculating and assigning \textit{chunks} of loop iterations to all the other entities (workers).
Exploiting this model has scaling~\cite{processorgroups}, and therefore, the DLS techniques have evolved to employ a hierarchical \mbox{master-worker} execution model~\cite{DHTSS}.
This model includes two levels of masters: global and local masters. 
The global master calculates and assigns chunks to local masters, each of them being responsible for calculating and assigning \mbox{sub-chunks} to its group of workers. 
The global master and local masters may exploit different DLS techniques.   
The use of DLS techniques at two levels is also referred to as a hierarchical DLS technique.

Hybrid \mbox{MPI+OpenMP} is a common programming approach to implement hierarchical DLS techniques. 
However, it has specific performance challenges, such as, the added overhead for the management of two levels of parallelism using two different runtime systems.

%
The OpenMP threads (workers) require synchronization before requesting and executing chunks, i.e., only the main thread is allowed to call MPI communication functions, such as MPI\_Send  and MPI\_Receive~\cite{HLS}. 
Otherwise, a complex implementation is needed to allow individual OpenMP threads to perform MPI calls.


This work proposes a novel approach for designing and developing hierarchical DLS techniques for \mbox{distributed-memory} systems.
It extends the distributed chunk calculation approach~\cite{RMAEleliemy} by allowing any group of workers to reside on a \mbox{shared-memory} system to form a shared work queue where the chunks to be executed by this group are stored. 
The novelty of the proposed approach lies in the fact that the responsibility of the work queue is shared among the workers of the group.
The present work considers the \mbox{shared-memory} features offered in the \mbox{MPI-3} standard to assign chunks to individual MPI processes in two stages. 
In the first stage, the fastest MPI process within a compute node obtains a chunk  based on a selected DLS technique.  
The fastest MPI process uses the obtained chunk to fill the local shared queue.
In the second stage, the MPI processes within the same \mbox{shared-memory} system use a different or a similar DLS technique to obtain \mbox{sub-chucks} from the shared local work queue.  

The importance of this work is enabling efficient and scalable implementations of hierarchical DLS techniques using a single programming model which promotes code clarity, maintainability, and decreased effort in debugging \mbox{work}. 
Moreover, it preserves the research efforts spent in developing DLS techniques using MPI. 
The main contributions of this work are as follows.
(1)~Proposal and implementation of a hierarchical version of DLS techniques for \mbox{distributed-memory} systems using the MPI+MPI approach.
(2)~Evaluation of using the \mbox{MPI+MPI} approach in developing hierarchical DLS techniques.

The remainder of this work is organized as follows. 
Section~\ref{sec:background} provides the necessary information to understand the DLS techniques, and presents similar research efforts reported in the literature. 
The proposed approach is described in Section~\ref{sec:proposed}. 
The experimental design and the evaluation results are discussed in Sections~\ref{sec:design} and \ref{sec:res}, respectively.
The conclusions and the future work are outlined in Section~\ref{sec:con}. 

%
%
%
\newpage
\section{Background and Related Work}
\label{sec:background}

\textbf{Load Balancing:}
The efficient execution of parallel applications requires balancing the workload among all processing elements, which means that all workers should finish almost at the same time.
In practice, such efficient executions are challenging to be achieved. 
Many factors, such as problem characteristics, algorithmic, and systemic variations hinder them.
Several load balancing techniques have been introduced in the literature~\cite{lb1,lb2, lb3, lb4, lb5, lb6} to fit the needs of different applications.

\textbf{Loop Scheduling Techniques:}
In scientific applications,  loops are the main source of parallelism~\cite{fang1990dynamic}. 
Loop scheduling techniques have been introduced to achieve a balanced load execution of loop iterations.
When loops have no \mbox{cross-iteration} dependencies, loop scheduling techniques map individual loop iterations to different processing elements aiming to have nearly equal finish times on all processing elements.    
Loop scheduling techniques can be categorized into static and dynamic loop scheduling. 
The time when scheduling decisions are taken is the key difference between both categories. 
Static loop scheduling~(SLS) techniques take scheduling decisions \textit{prior to applications' execution}, while dynamic loop scheduling~(DLS) techniques take scheduling decisions \textit{during  applications' execution}.
Therefore, SLS techniques have less scheduling overhead than DLS, and  DLS techniques can achieve better load balanced executions than SLS in \textit{highly dynamic execution} environments.  
 
 DLS techniques can further be divided into \mbox{non-adaptive} and adaptive techniques. 
 The \mbox{non-adaptive} techniques utilize certain information that is obtained prior to the application execution.
 The adaptive techniques regularly obtain information during the application execution, and the scheduling decisions are taken based on that new information.  
The adaptive techniques incur a significant scheduling overhead compared to \mbox{non-adaptive} techniques, and outperform the \mbox{non-adaptive} ones in \emph{highly irregular execution} environments.
 
 \textbf{Selected DLS Techniques:}
 In addition to the extreme cases of loop scheduling techniques: \mbox{fully-static} (STATIC) and \mbox{fully-dynamic} (SS), we selected three DLS techniques: guided \mbox{self-scheduling}~(GSS)~\cite{GSS}, trapezoid \mbox{self-scheduling}~(TSS)~\cite{TSS}, and Factoring~(FAC)~\cite{FAC}.
 These techniques are remarkable points on the DLS spectrum, since they have competitive performance in different applications, and they are at the basis of other DLS techniques, such as weighted factoring~\cite{WF}, adaptive weighted factoring~(AWF)~\cite{AWF}, and trapezoid factoring \mbox{self-scheduling}~(TFSS)~\cite{TFSS}. 
 
In STATIC, each processing element obtains a chunk of~$N/P$, where~$N$ and~$P$ are the total number of iterations and the total number of processing elements, respectively.
STATIC incurs the lowest scheduling overhead.
 
 SS~\cite{SS} is a dynamic \mbox{self-scheduling} technique where the chunk size is always one iteration.
 SS has the highest scheduling overhead. 
 However, SS can achieve a highly \mbox{load-balanced} execution in highly irregular execution environments.
 
In GSS~\cite{GSS} at every scheduling step, the remaining loop iterations are divided by the total number of processing elements. 
GSS is a compromise between the maximum load balancing that can be achieved using SS and the lowest scheduling overhead that is incurred in STATIC.   
   
 Unlike GSS, TSS~\cite{TSS} uses a linear function to decrement chunk sizes. 
This linearity results in low scheduling overhead in each scheduling step compared to GSS.

FAC~\cite{FAC} is another DLS technique that schedules the loop iterations in batches of \mbox{equally-sized} chunks.
FAC evolved from comprehensive probabilistic analyses, and assumes prior knowledge about the standard deviation of loop iterations’ execution times $\sigma$ and their mean execution time~$\mu$.
Another practical implementation of FAC, denoted FAC2, assigns half of the remaining loop iterations for every batch. 
The initial chunk size of FAC2 is half of the initial chunk size of GSS.
If more \mbox{time-consuming} loop iterations are at the beginning of the loop, FAC2 may balance their execution better than GSS.
 
 \textbf{Related Work:}
Most of the DLS techniques were introduced between the late of 1980s and the early of 2000s~\cite{SS,GSS,TSS,FAC,WF,AWF,AWFBC,PLS,HLS}.
The DLS techniques target loop executions on both, \mbox{shared-memory} and \mbox{distributed-memory} systems. 
Therefore, their implementation strategies and technologies evolved to fit both system types.

On \mbox{shared-memory} systems, DLS techniques have been implemented using \mbox{multi-threading}. 
Threads share certain variables (the last scheduled loop index and the remaining loop iterations) that control the loop execution, and each thread calculates its own chunk size based on the values of these variables. 
Atomic operations or locks are used to keep the shared variables up to date.
Open multi-processing (OpenMP) is considered as one of the most dominant \mbox{multi-threading} APIs for \mbox{shared-memory} systems.
The \mbox{OpenMP-5} standard~\cite{OpenMP} uses the \texttt{schedule} clause to define three different loop scheduling options: \textit{static}, \textit{dynamic}, and \textit{guided}.
Table~\ref{tab:mapping} shows the mapping between the OpenMP schedule clause and the DLS techniques discussed earlier in this section.
\begin{table}[!b]
	\centering
		\caption{Mapping between the DLS techniques and the OpenMP schedule clause options}
		\label{tab:mapping}
	\begin{tabular}{l|l}
		DLS technique & OpenMP schedule clause \\ \hline
		STATIC        & schedule(static)       \\
		SS            & schedule(dynamic,1)    \\
		GSS           & schedule(guided,1) 
	\end{tabular}
\end{table}
The OpenMP standard uses the \textit{schedule(runtime)} clause to allow users to select a certain scheduling technique during the execution of the application.
A recent research effort implemented other DLS techniques, namely TSS, FAC2, WF, and random \mbox{self-scheduling} in an open source OpenMP runtime  library, called~\mbox{LaPeSD-libGOMP}\footnote{https://github.com/lapesd/libgomp}~\cite{Ciorba:2018}. 
The authors exploited the \textit{schedule(runtime)} clause together with certain \mbox{user-defined} environment variables to select one of the DLS techniques that they implemented in the \mbox{LaPeSD-libGOMP} runtime.

On \mbox{distributed-memory} systems, the dominant approach for developing DLS techniques is the message-passing interface~(MPI)~\cite{MPIForum}. 
The \mbox{MPI-based} implementation requires employing a \mbox{master-worker} execution model, where a certain MPI process, called master is responsible for calculating and assigning chunks to other free and requesting MPI processes, called workers.

A load balancing tool, called~\mbox{DLB\_tool} that integrates several DLS techniques was introduced~\cite{LBtool}.
The \mbox{DLB\_tool} employed the \mbox{master-worker} execution model implemented using the classical \mbox{two-sided} MPI communication.
The tool assumed a \mbox{non-dedicated} master process that participates in the loop execution similar to the workers.

An enhanced tool for dynamic load balancing, called  dynamic load balancing library~(DLBL)~\cite{DLBLtool} was the first to utilize MPI \mbox{one-sided} communication.
In DLBL, the master process receives work requests, and for each request, it calculates the size of the new chunk.
The master process calls a handler function on the worker side. 
Workers obtain the data of the new chunks from the master process.

In the \mbox{master-worker} execution model, the number of requests that could be received by the master process is proportional to the total number of workers.
For a large number of workers, the master may simultaneously receive a large number work requests, and if the handling og the work requests is inefficient, the master becomes a performance bottleneck.  
To overcome such limitation, certain research efforts proposed the use of the hierarchical \mbox{master-worker} approach. 
For instance, a distributed self-scheduling scheme~(HDSS) using the hierarchical \mbox{master-worker} model was introduced~\cite{DHTSS}.
Unlike the \mbox{LB-tool}, HDSS dedicated the master process for handling the worker requests. 
The proposed scheme was implemented using MPI and its classical \mbox{two-sided} communication.

Another research effort discussed the adverse impact of the \mbox{foreman-worker} (master-worker) model~\cite{processorgroups} on DLS techniques.
The authors suggested a new execution model using processor groups. 
The idea was to form a few groups of processors, where each group executes a specified portion of the iteration space using the \mbox{master-worker} model.
However, the master process of each group has to periodically update a global master process, called manager, with  the ratio of the remaining iterations and the available workers.
When the reported ratio exceeds a certain threshold, the manager migrates workers between processor groups.
This research effort is similar to the present work as both balance the loop execution using two levels of scheduling.
However, the present work differs by exploiting different DLS techniques at the first level while the suggested execution model in~\cite{processorgroups} statically divides the loop iteration space among processor groups.
Moreover, the present work avoids worker migration that may result in performance degradation when the cost of managing and serving requests from the migrated workers become relatively large. 

Recently, a distributed \mbox{chunk-calculation} approach was proposed for developing DLS techniques executing on \mbox{distributed-memory} systems~\cite{RMAEleliemy}. 
This approach eliminated the use of the \mbox{master-worker} model by exploiting the \mbox{one-sided} communication features offered in the \mbox{MPI-3} standard.   
The distributed \mbox{chunk-calculation} approach uses formulas of DLS techniques that depend on a single value, namely, the latest scheduling step.
The main idea is that each worker can atomically get and increment the latest scheduling step, then workers can simultaneously and independently compute their own chunk sizes.
This avoids the need of having a master entity that sequentially assigns and computes chunks to requesting workers. 

Modern high performance computing~(HPC) systems are clusters of multi- and \mbox{many-core} systems connected via \mbox{high-speed} interconnection networks~\cite{MPIOpenMP3}.
Developers often combine two different programming models to target such systems.
A common approach is to use MPI~\cite{MPIForum} for inter-node communication and OpenMP~\cite{OpenMP} for programming the \mbox{shared-memory} systems~\cite{MPIOpenMP2}.
In the context of DLS techniques, the hierarchical loop scheduling~(HLS)~\cite{HLS} was one of the earliest efforts to use MPI+OpenMP programming model.
In HLS, a free worker (MPI process) requests a chunk from the master rank which calculates and assigns the chunk based on a certain performance function~\cite{PLS}.
The  workers (MPI processes) locally use OpenMP loop scheduling techniques, such as static, dynamic, and guided to execute the assigned chunk.

The current work addresses the following limitations of the HLS approach: (1)~the implicit synchronization between all OpenMP threads at the end of the execution of each chunk and (2)~the limited choice of DLS techniques currently specified in the OpenMP standard.

\section{The Hierarchical DLS Approach}
\label{sec:proposed}

\begin{figure}[!t]
	\centering
	\includegraphics[width=\columnwidth]{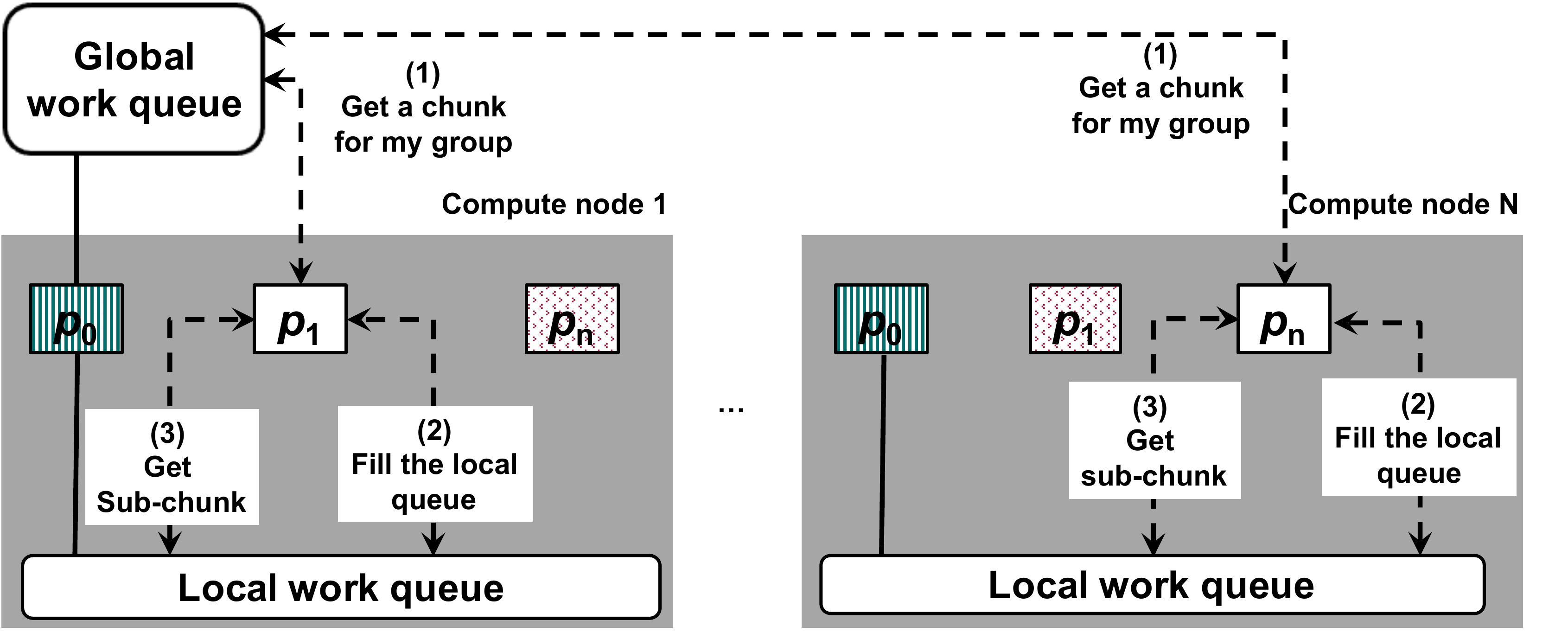}\\
	\includegraphics[width=\columnwidth]{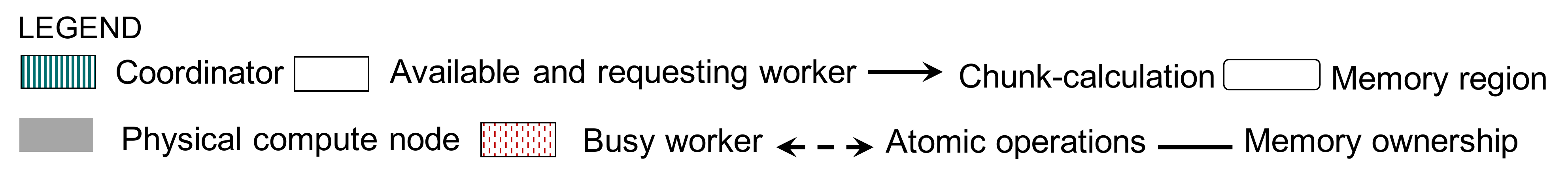}
	\caption{Illustration of the proposed hierarchical DLS techniques using MPI+MPI-based approach. }
	\label{fig:proposed}
\end{figure}

The proposed approach applies two DLS techniques at the intra- and \mbox{inter-node} levels as follows: 
one MPI process creates a global \mbox{shared-memory} region, called \emph{global work queue}.
This global queue stores information regarding the latest scheduling step and the total scheduled loop iterations~\cite{RMAEleliemy}.
Using \texttt{MPI\_Win\_allocate\_shared}, the MPI processes within one compute node create another \mbox{shared-memory} region, called \emph{local work queue}. 
This local queue stores information regarding the latest scheduling step and the total scheduled loop iterations by the MPI processes within that physical node.
Whenever an MPI process becomes free, it obtains a \mbox{sub-chunk} from the local work queue.
If there are no \mbox{sub-chunks}, the MPI process tries to obtain a chunk from the global work queue, and fills the empty local work queue. 
In the proposed hierarchical DLS approach, the MPI processes do not wait for each other to fill the local queue. 
The responsibility of obtaining work is not assigned to a specific MPI process, as the fastest MPI process always takes this responsibility. 
 Figure~\ref{fig:proposed} illustrates the proposed hierarchical DLS approach and its \mbox{MPI+MPI}-based approach.

\begin{figure}[!b]
	\centering
	\includegraphics[width=\columnwidth]{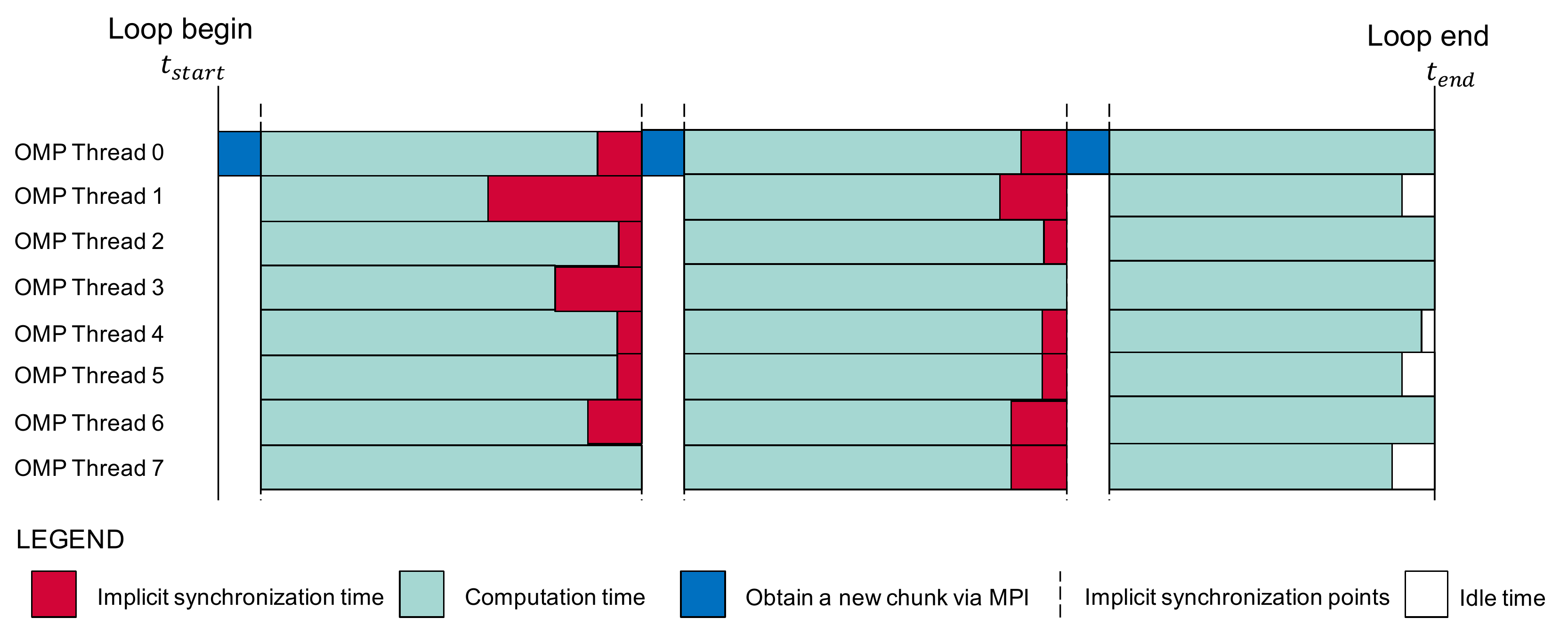}
	\caption{Illustration of the undesired synchronization with the MPI+OpenMP approach at the \mbox{shared-memory} level.}
	\label{fig:openmp}
\end{figure}

Unlike existing MPI+OpenMP implementations of DLS techniques, the proposed approach avoids the implicit synchronization that is required at the end of executing the chunks.
Figure~\ref{fig:openmp} illustrates the undesired implicit thread synchronization when using the MPI+OpenMP approach.
OpenMP Thread~1 finished its \mbox{sub-chunk} earlier than the rest of the threads.
However, it has to wait for the slowest OpenMP Thread (OpenMP Thread~7).
At the second chunk, the same scenario was repeated when OpenMP Threads~6 and~7 finished earlier than the rest. 

\begin{figure}[!b]
	\centering
	\includegraphics[width=\columnwidth]{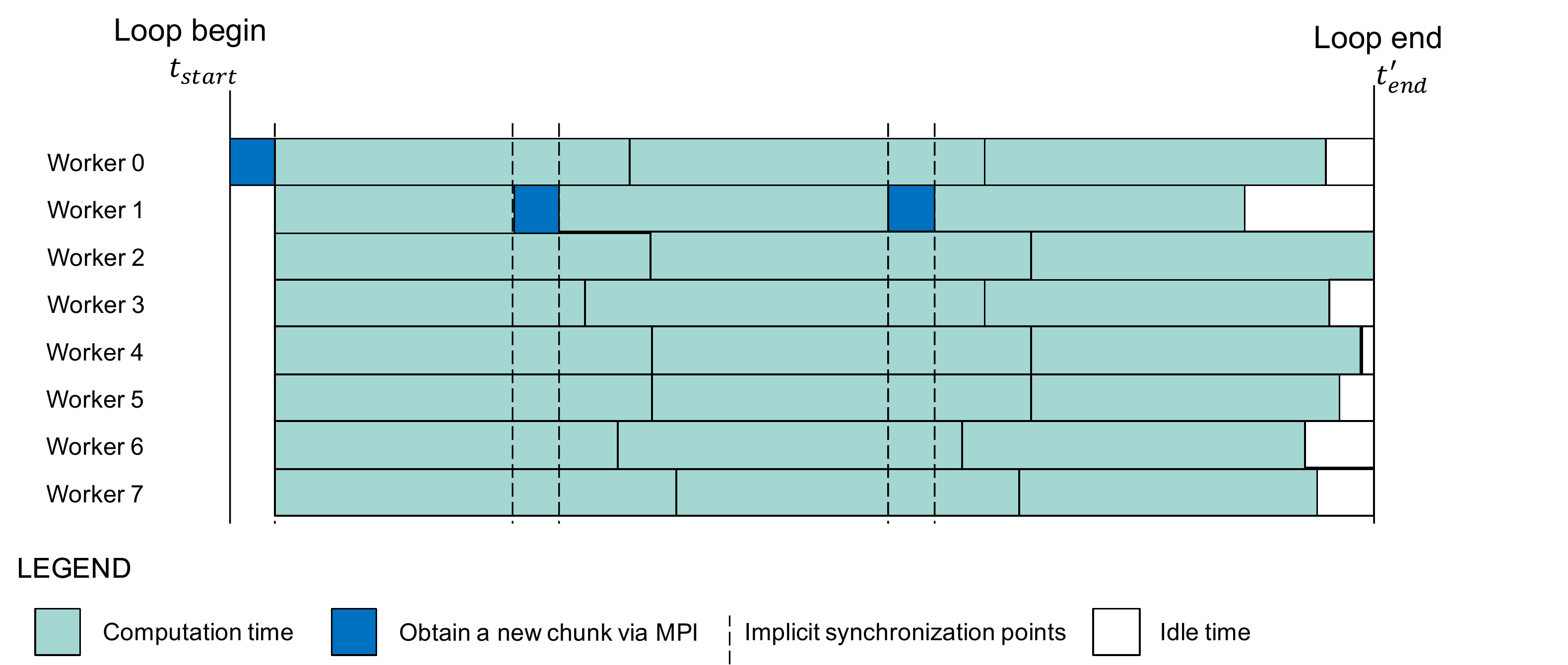}
	\caption{Illustration of an ideal execution scenario at the \mbox{shared-memory} level. }
	\label{fig:ideal}
\end{figure}

Figure~\ref{fig:ideal} shows the desired optimal execution scenario at the \mbox{shared-memory} level.
Worker~1 finished earlier than the rest, however, it immediately obtained a new chunk to fill the local queue, and then it obtained a \mbox{sub-chunk} for itself.
Once any other worker finished its \mbox{sub-chunk}, it could directly obtain a \mbox{sub-chunk} from the most recent chunk obtained by Worker~1. 
In Figure~\ref{fig:ideal}, the parallel time to execute the loop $t^\prime_{end}$ is less than $t_{end}$ in Figure~\ref{fig:openmp}.

Compared to the proposed approach, one could state that the main issue of the \mbox{MPI+OpenMP} approach is the \emph{implicit barrier} at the end of executing each chunk of a loop iteration.
Such an issue could be solved by using  the \emph{nowait} clause that allows OpenMP threads to continue their execution when there are no more loop iterations to execute.
However, the use of the \emph{nowait} clause requires all OpenMP threads to initiate \texttt{MPI\_Send} and \texttt{MPI\_Recv} calls, i.e., the fastest OpenMP thread may differ from one chunk to another.
Therefore, the implementation would  require many synchronization statements to guarantee the exclusive request of new chunks for only one thread at a time.
This leads to more complicated codes, which are hard to tune and maintain.  


Another significant advantage  of the \mbox{MPI+OpenMP} approach is the flexible selection of different loop scheduling technique; i.e., one may specify the scheduling technique using the \texttt{schedule} clause~\cite{Ciorba:2018}. 
The \mbox{MPI-based} libraries that implement DLS techniques, such as the \mbox{DLB\_tool}~\cite{DLBLtool}, often guarantee the same flexibility, e.g.,
one input parameter specifies the selected DLS technique.
The present work discusses and evaluates the advantages and disadvantages of the proposed approach.
However, introducing a library that uses the proposed approach and provides the same flexibility is feasible and planned as future work.

%
%



\section{Experimental Setup}
\label{sec:design}
\textbf{Selected Applications:} Two scientific applications are used to assess and compare the performance of the proposed \mbox{MPI+MPI} hierarchical DLS approach: PSIA~\cite{psia,psia2} and Mandelbrot~\cite{mandelbrot1980fractal}. 
PSIA is a parallel version of the \mbox{spin-image} algorithm~(SIA), which converts a 3D object into a set of 2D images~\cite{sia}. 
The generated \mbox{spin-images} can be used as a shape descriptor. 
The efficient generation of \mbox{spin-images} has a \mbox{high-interest} for 3D object recognition and categorization.
Dynamic loop scheduling techniques have been used to implement an efficient parallel \mbox{spin-image} generation algorithm~\cite{Eleliemy:2017b}.
Therefore, the present work examines the potential of using hierarchical DLS techniques in enhancing the performance of PSIA. 

Mandelbrot is a parallel application that calculates the Mandelbrot set~\cite{mandelbrot1980fractal}. 
Mandelbrot is selected as a test application due to high algorithmic load imbalance that motivated its use as a kernel for DLS performance evaluation in the literature~\cite{DHTSS,HLS,PLS}.	
Both applications are \mbox{computationally-intensive} and contain a single large parallel loop that dominates the application execution time.

\textbf{Implementation Approaches:}
Hierarchical DLS techniques can be implemented either using the hierarchical \mbox{master-worker}~\cite{DHTSS} or using the distributed \mbox{chunk-calculation} model~\cite{RMAEleliemy}.
The present work evaluates the use of two different implementations, \mbox{MPI+OpenMP} and \mbox{MPI+MPI}, to complement \emph{the distributed \mbox{chunk-calculation} approach} (see Section~\ref{sec:background}).

The \mbox{MPI+OpenMP} implementation complements the distributed \mbox{chunk-calculation} approach by the use of OpenMP at the \mbox{shared-memory} level.
It maps one MPI process per each compute node.
The mapped MPI processes communicate and cooperate to obtain  chunks using one of the following DLS techniques: STATIC, SS, GSS, TSS, and FAC2. 
Every MPI process uses the OpenMP runtime to create a number of threads equal to the number of its computing cores. 
The threads use the OpenMP loop scheduling techniques (static, dynamic, and guided) to execute the chunks obtained from their (owner) MPI process.
The \mbox{MPI+MPI} implementation complements the \mbox{distributed-chunk} calculation approach by the use of  MPI \mbox{shared-memory} capabilities as explained in Section~\ref{sec:proposed}, i.e., it forms shared local queues at the compute node level (see Figure~\ref{fig:proposed}).  

\textbf{Selected  System:}
The target hardware is a small HPC cluster (called miniHPC~\cite{miniHPC}) that consists of 26~compute nodes.
The first 22 nodes are two-socket Intel Xeon \mbox{E5-2640} processors with 20~cores, 64~GB RAM, and 2.4 GHz~CPU frequency.
The remaining 4 nodes are standalone Intel Xeon Phi~7210 manycore processors with~64 cores and 96~GB~RAM.
All nodes are interconnected using Intel \mbox{Omni-Path} fabric in a \mbox{non-blocking} \mbox{fat-tree topology}. 
The network bandwidth and latency are 100 GBit/s  and 100 ns, respectively.
The miniHPC cluster is used for educational and research purposes. 
Therefore, sixteen identical nodes (representing 70\% of the computational power of this cluster) were dedicated to the experimental evaluation presented in this work.
All parallel computing nodes use CentOS Linux release 7.5.1804 as operating system.
Slurm 17.02.7 is configured to be the batch system of the cluster.
The Intel MPI version 18.0.3 was used to compile and the O3 compilation flag was enabled.

\newpage
\section{Experimental Evaluation}
\label{sec:res}

The OpenMP standard currently supports three loop scheduling techniques: static, dynamic, and guided~(cf. Table~\ref{tab:mapping}).
As discussed in Section~\ref{sec:background}, more loop scheduling techniques were implemented in an OpenMP runtime library, called \mbox{LaPeSD-libGOMP}~\cite{Ciorba:2018}.
However, for accurate performance measurements, we wanted to use the most optimized software  installed on miniHPC.
Given that miniHPC (the target system) is an \mbox{Intel-based} cluster, the Intel software stack was selected, and therefore, scheduling experiments that have TSS and FAC at the \mbox{shared-memory} level were only performed using the proposed \mbox{MPI+MPI} approach.
The use of \mbox{LaPeSD-libGOMP}, instead of Intel OpenMP runtime library, enables more DLS techniques, and it is planned as future work.

In this section, the notation of X+Y is used to represent scheduling combinations, where X is a DLS technique used at the \mbox{inter-node} level and Y is a DLS technique used to at the \mbox{intra-node} level. 
X and Y can refer to one DLS technique.

\begin{figure*}
	\captionsetup[subfigure]{justification=centering}
	\centering
	\begin{subfigure}{\textwidth}
		\includegraphics[width=\textwidth]{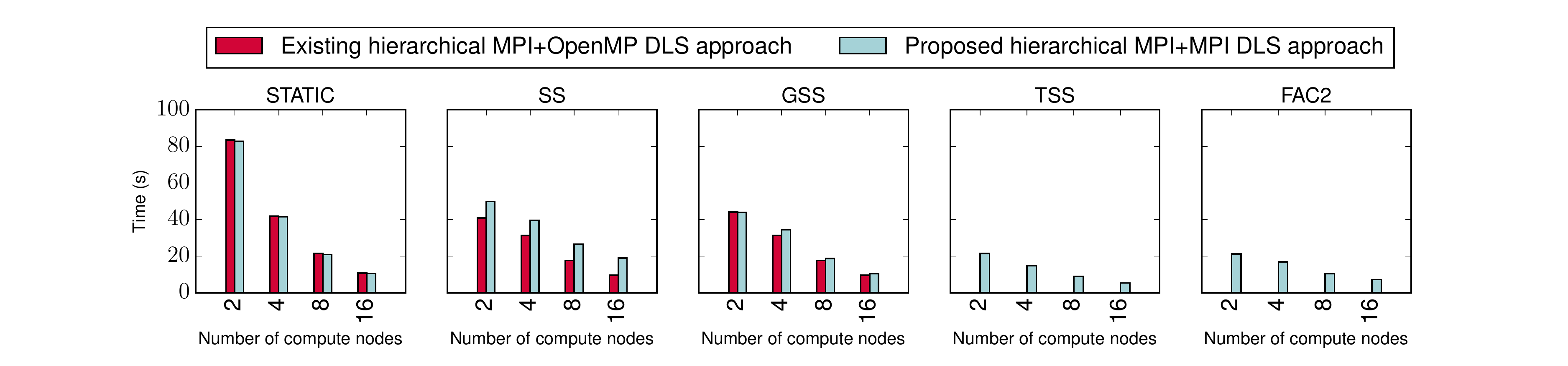}
		\caption{Mandelbrot}
		\label{fig:mandel_static}
	\end{subfigure}
	\centering
	\begin{subfigure}{\textwidth}
		\includegraphics[clip, trim=0cm 0cm 0cm 2.1cm, width=\textwidth]{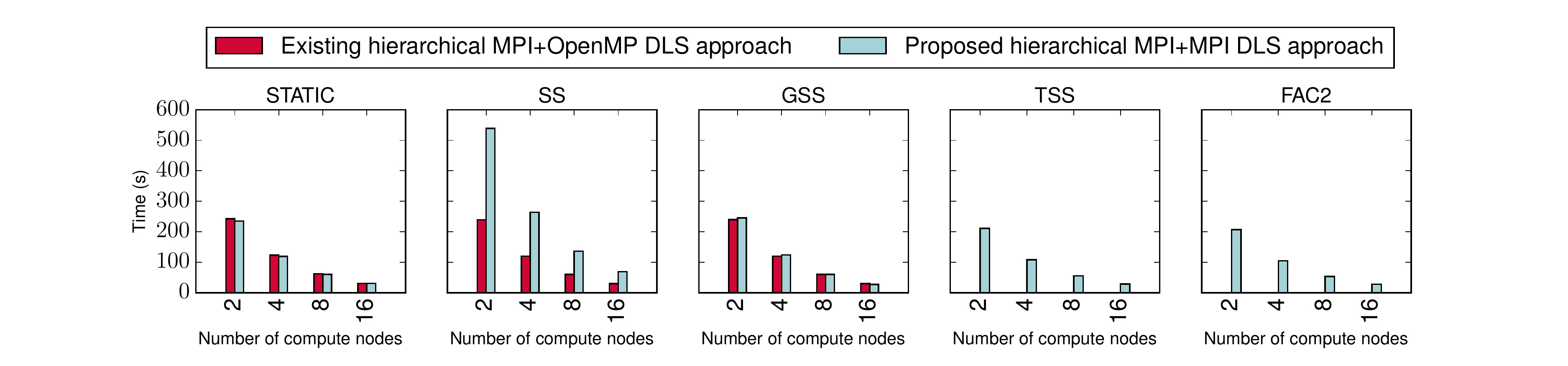}
		\caption{PSIA}
		\label{fig:pisa_static}
	\end{subfigure}
	\caption{Parallel execution time of the main loop of both applications, Mandelbrot and PSIA.
		For the MPI+OpenMP approach, each worker is an OpenMP thread, and the total MPI processes per one compute node is one process. For the MPI+MPI approach, each worker is an MPI process, and the total MPI processes per one compute node is 16 processes. STATIC is the first level of scheduling (the inter-node scheduling).}
	\label{fig:static}
\end{figure*}

\begin{figure*}[!b]
	\captionsetup[subfigure]{justification=centering}
	\centering
	\begin{subfigure}{\textwidth}
		\includegraphics[width=\textwidth]{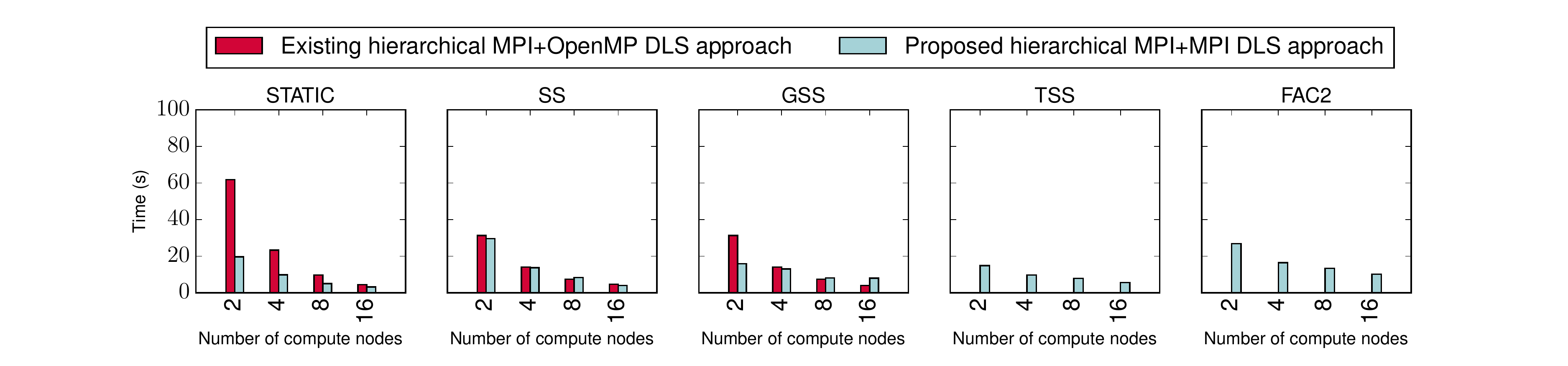}
		\caption{Mandelbrot}
		\label{fig:mandel_gss}
	\end{subfigure}
	\centering
	\begin{subfigure}{\textwidth}
		\includegraphics[clip, trim=0cm 0cm 0cm 2.1cm, width=\textwidth]{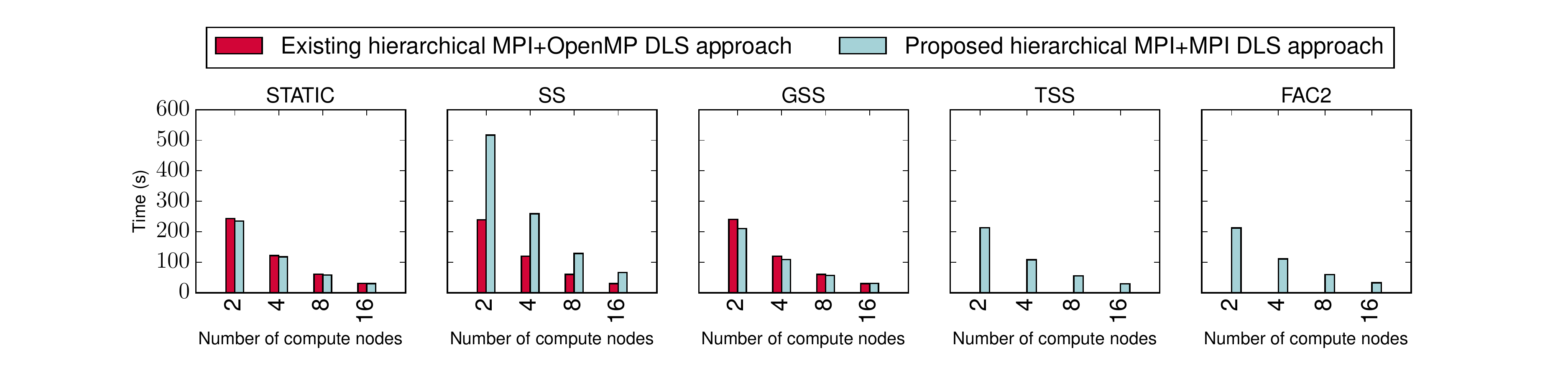}
		\caption{PSIA}
		\label{fig:pisa_gss}
	\end{subfigure}
	\caption{Parallel execution time of the main loop of both applications, Mandelbrot and PSIA.
		For the MPI+OpenMP approach, each worker is an OpenMP thread, and the total MPI processes per one compute node is one process. For the MPI+MPI approach, each worker is an MPI process, and the total MPI processes per one compute node is 16 processes. GSS is the first level of scheduling (the inter-node scheduling).}
	\label{fig:gss}
\end{figure*}

\begin{figure*}[!b]
	\captionsetup[subfigure]{justification=centering}
	\centering
	\begin{subfigure}{\textwidth}
		\includegraphics[width=\textwidth]{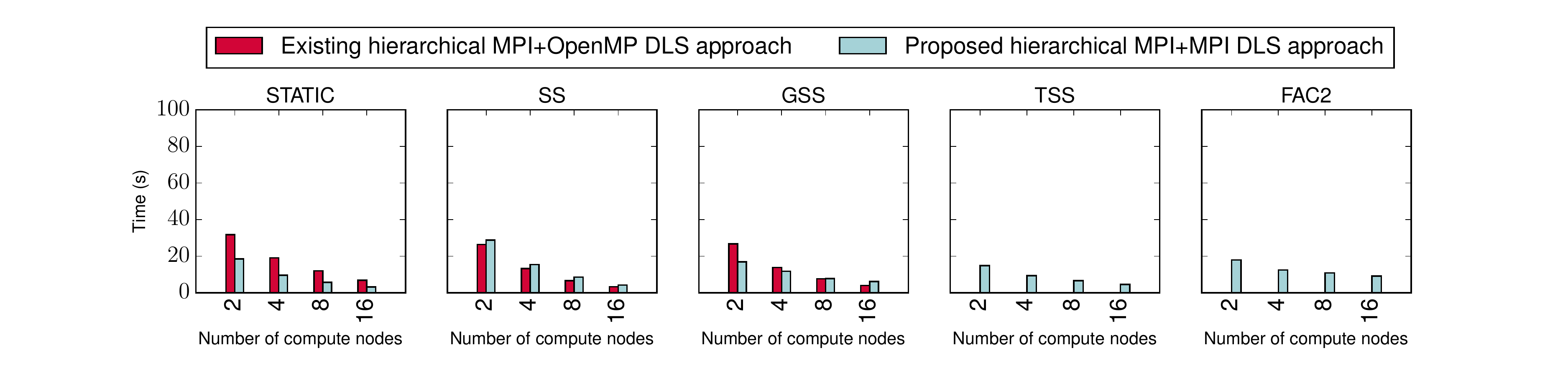}
		\caption{Mandelbrot}
		\label{fig:mandel_tss}
	\end{subfigure}
	\centering
	\begin{subfigure}{\textwidth}
		\includegraphics[clip, trim=0cm 0cm 0cm 2.1cm, width=\textwidth]{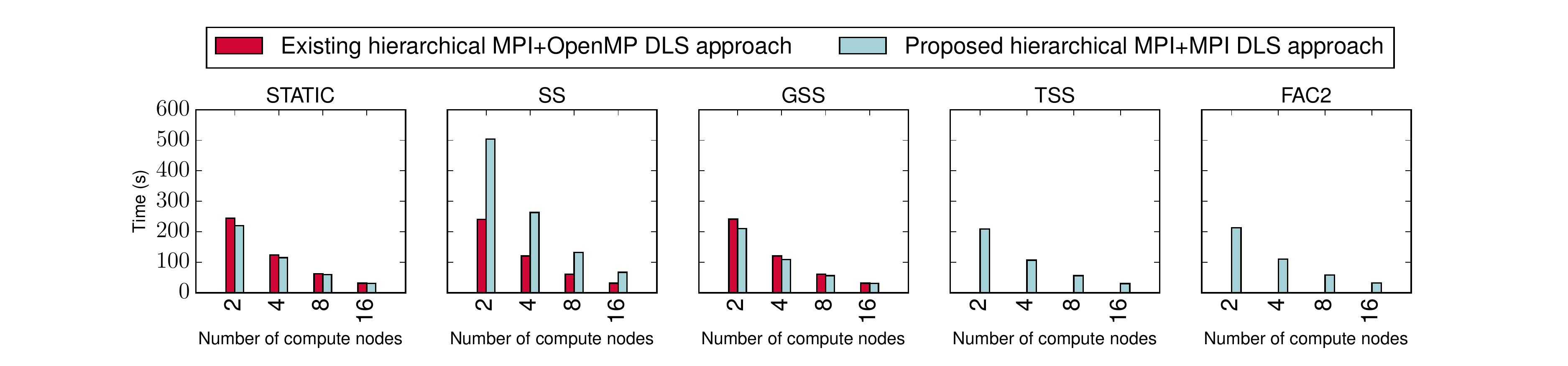}
		\caption{PSIA}
		\label{fig:pisa_tss}
	\end{subfigure}
	\caption{Parallel execution time of the main loop of both applications, Mandelbrot and PSIA.
		For the MPI+OpenMP approach, each worker is an OpenMP thread, and the total MPI processes per one compute node is one process. For the MPI+MPI approach, each worker is an MPI process, and the total MPI processes per one compute node is 16 processes. TSS is the first level of scheduling (the inter-node scheduling).}
	\label{fig:tss}
\end{figure*}

\begin{figure*}
	\captionsetup[subfigure]{justification=centering}
	\centering
	\begin{subfigure}{\textwidth}
		\includegraphics[width=\textwidth]{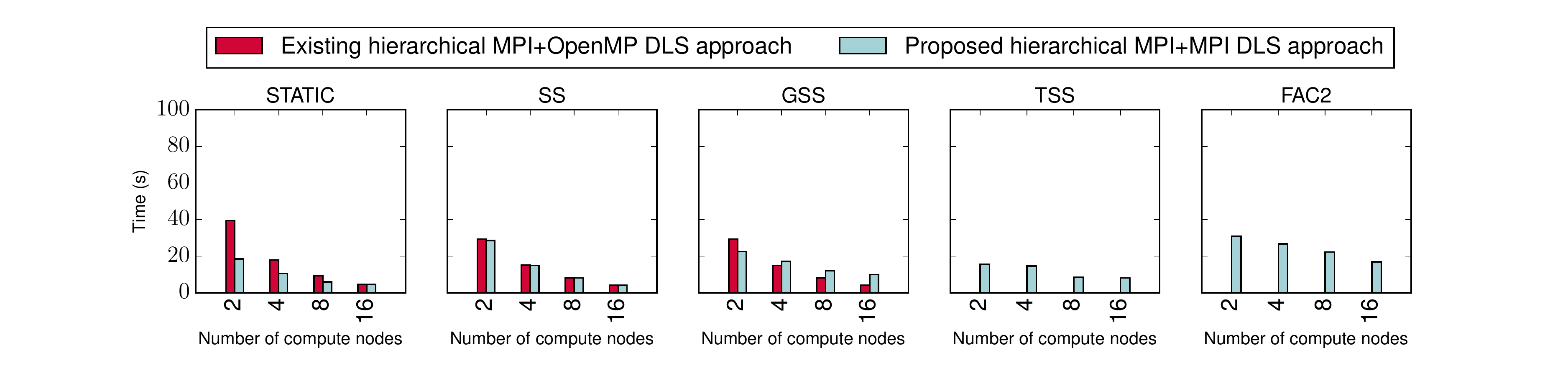}
		\caption{Mandelbrot}
		\label{fig:mandel_fac}
	\end{subfigure}
	\centering
	\begin{subfigure}{\textwidth}
		\includegraphics[clip, trim=0cm 0cm 0cm 2.1cm, width=\textwidth]{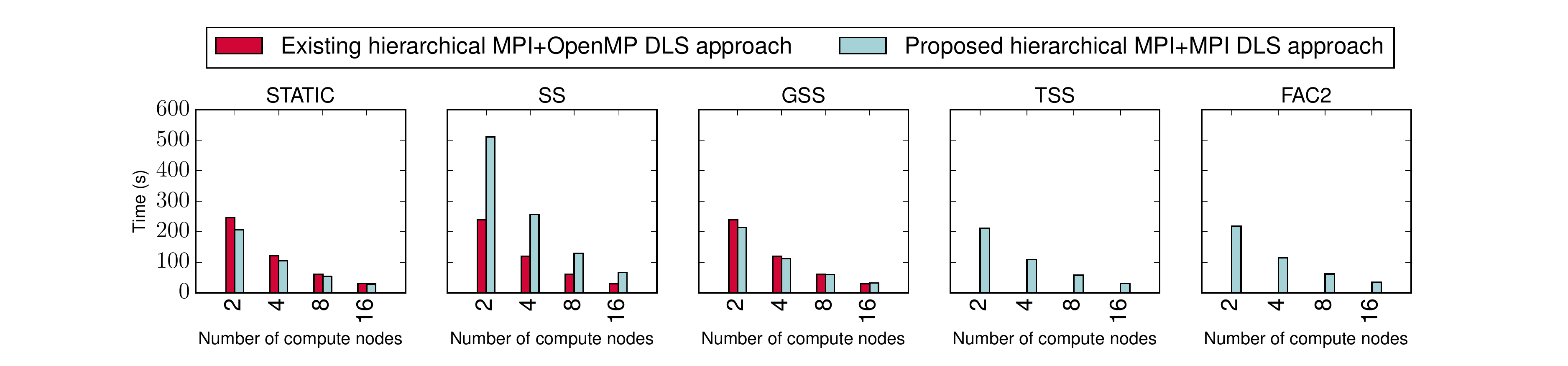}
		\caption{PSIA}
		\label{fig:pisa_fac}
	\end{subfigure}
	\caption{Parallel execution time of the main loop of both applications, Mandelbrot and PSIA.
		For the MPI+OpenMP approach, each worker is an OpenMP thread, and the total MPI processes per one compute node is one process. For the MPI+MPI approach, each worker is an MPI process, and the total MPI processes per one compute node is 16 processes. FAC2 is the first level of scheduling (the inter-node scheduling).}
	\label{fig:fac}
\end{figure*}

Figures~\ref{fig:static} to~\ref{fig:fac} show the performance of executing Mandelbrot and PSIA with two levels of DLS techniques.
Figure~\ref{fig:static} shows the first combination of DLS techniques where STATIC is used to schedule the workload across multiple compute nodes.
An important observation is that when SS is selected to schedule the workload within one computing node, the proposed \mbox{MPI+MPI} approach has the poorest performance compared to the \mbox{MPI+OpenMP}.
The reason  is due to the use of \texttt{MPI\_Win\_lock} and \texttt{MPI\_Win\_sync}. 
These functions provide an exclusive access to the local work queue (see Figure~\ref{fig:proposed}), and consequently, maintain the work queue.
The \texttt{MPI\_Win\_lock} uses a lock polling technique where an MPI process repeatedly issues \mbox{lock-attempt} messages until the lock is granted~\cite{zhao2016scalability}.
Consequently, the number of \mbox{lock-attempt} messages increases when multiple processes try to acquire the same lock at the same time, and more overhead is introduced.

Another observation is that all hierarchical DLS techniques, except SS, implemented with the proposed \mbox{MPI+MPI} approach have the same performance compared to their counterparts implemented using the MPI+OpenMP approach.
The reason is that using STATIC at the \mbox{inter-node} level means there is only one scheduling round at that level.
However, achieving the same results also indicates that the proposed approach did not introduce a significant overhead to the DLS techniques.

Figure~\ref{fig:gss} shows the second combination of DLS techniques where GSS is used to schedule the workload across the compute nodes.
For both applications, the proposed \mbox{MPI+MPI} approach outperformed the \mbox{MPI+OpenMP}. 
The results of the \mbox{GSS+STATIC} combination shows the advantage of the proposed approach, where avoiding the unnecessary synchronization between the workers (OpenMP threads) has a significant adverse impact.
For instance, in Mandelbrot and using the proposed approach, the parallel execution times of the \mbox{GSS+STATIC} combination were 19.6 and 3.1 seconds on the smallest and the largest system sizes, respectively.
The same scheduling combination \mbox{GSS+STATIC} using MPI+OpenMP required 61.5 and 4.5 seconds on the smallest and the largest system sizes, respectively.
In PSIA, the performance trend was repeated, i.e., the \mbox{GSS+STATIC} using the proposed MPI+MPI approach outperformed its counterpart implemented using the MPI+OpenMP approach. 
For instance, on the smallest systems size, the parallel execution times were 233 and 245 seconds using the proposed MPI+MPI approach and the MPI+OpenMP approach, respectively.
However, the two approaches had the same performance when executing on the largest system size.
The reason is the decreased load imbalance in PSIA compared to that in Mandelbrot.
For the \mbox{GSS+GSS} combination, the DLS techniques implemented using the proposed MPI+MPI approach also outperformed their counterparts implemented using the MPI+OpenMP approach.

	As discussed in earlier in this Section, we decided to use the Intel software stack.
	Therefore, it was not possible to perform the remaining combinations: \mbox{GSS+TSS} and \mbox{GSS+FAC2} using MPI+OpenMP, i.e., the Intel OpenMP runtime library only supports the following loop scheduling techniques: static, dynamic and guided. 


Figures~\ref{fig:tss} and~\ref{fig:fac} show the third and the fourth combinations of DLS techniques where TSS and FAC2 are used to schedule the workload across multiple compute nodes, respectively.
Similar to Figure~\ref{fig:gss}, the proposed approach significantly outperformed the \mbox{MPI+OpenMP} approach when STATIC is selected for scheduling the computational workload within one compute node. 
For the rest of the scheduling combinations, both approaches have the same performance. 
The only exception is when applying SS at the \mbox{shared-memory} level. 
The proposed approach has the worst performance compared to the MPI+OpenMP approach. 
The reason is that SS achieves the maximum load balance, and most of the workers (OpenMP threads) finishes at the same time. 
This scenario will avoid the long synchronization time before getting new chunks.

The last observation is related to the performance of the PSIA when applying any combination that has the SS using the proposed approach.
PSIA has less load imbalance than Mandelbrot and the proposed approach has already a significant overhead when employing SS, and consequently, the adverse impact of the large associated scheduling overhead of SS is more visible in PSIA than Mandelbrot.

\newpage
\section{Conclusion and Future Work}
\label{sec:con}
The implementation of hierarchical dynamic loop self-scheduling techniques is essential to enable scalable application performance.
When STATIC is used for the \mbox{intra-node} level scheduling, the implementation that uses  \mbox{MPI+MPI}  approach outperformed the one that uses hybrid \mbox{MPI+OpenMP} approach.
This highlights and confirms the capability of the MPI+MPI approach to eliminate the unrequired synchronization at the \mbox{intra-node} level. 
On the contrary, the \mbox{MPI+MPI} approach shows significant limited performance when many MPI processes on the same \mbox{shared-memory} system try to simultaneously access the local work queue.
The important observation of the present work is that the scheduling overhead associated with using MPI \mbox{shared-memory} to implement DLS techniques is higher than OpenMP. 
Therefore, the use of the MPI+MPI approach is only recommended for developing hierarchical DLS techniques when its associated overhead is less than the synchronization overhead associated with the use of OpenMP. 
The assessment of the performance and the development efforts associated with the use of \textit{nowait} clause compared to the use of the proposed \mbox{MPI+MPI} hierarchical DLS approach are planned as future work.
Moreover, a detailed evaluation of the synchronization overhead between the OpenMP threads with and without the use of \textit{nowait} clause is needed and planned as future work.

\section*{Acknowledgment}
This work has been supported by the Swiss National Science Foundation in the context of the Multi-level Scheduling in Large Scale High Performance Computers” (MLS) grant number 169123 and by the Swiss Platform for Advanced Scientific Computing (PASC) project SPH-EXA: Optimizing Smooth Particle Hydrodynamics for Exascale Computing. 
\newpage


 \bibliographystyle{IEEEtran}
 \balance
 \bibliography{references}

\end{document}